\documentstyle[psfig,preprint,aps]{revtex}
\begin{document}

\draft
\title{On the Determination of Small $x$ Shadowing Corrections at HERA}
\author{
 A. L. Ayala Filho $^{1,\dagger}$,\footnotetext{$^{\dagger}$E-mail:ayala@ufpel.tche.br}  M. B. Gay  Ducati $^{2,*}$\footnotetext{$^{*}$E-mail:gay@if.ufrgs.br}
 and
Victor P.  Gon\c{c}alves $^{2,**}$\footnotetext{$^{**}$E-mail:barros@if.ufrgs.br}} 
\address{$^1$ Instituto de F\'{\i}sica e Matem\'atica, Universidade Federal de Pelotas \\
Caixa Postal 354, CEP 96010-090, Pelotas, RS, BRAZIL \\
$^2$ Instituto de F\'{\i}sica, Universidade Federal
do Rio Grande do Sul\\ Caixa Postal 15051, CEP 91501-970, Porto Alegre, RS, BRAZIL}

\maketitle
\begin{abstract}
The recent $F_2$ data can be described using the DGLAP evolution equations with
an appropriate choice of input distributions and the choice of the starting 
scale for the $Q^2$ evolution. We demonstrate in this paper that we cannot 
conclude that there are no significant shadowing corrections at HERA kinematic 
region using  $F_2$ data only.
In this paper we calculate the shadowing corrections to the longitudinal 
structure function $F_L(x,Q^2)$ and to the charm component of the proton 
structure function $F_2^c(x,Q^2)$ at the HERA kinematic region using an 
eikonal approach. 
We demonstrate that the  shadowing corrections to these observables are 
very large and that the charm production is strongly modified at small-$x$. 
Our results agree with the recent few H1 data.
\end{abstract}


\bigskip







\section{Introduction}

The parton high density regime in deep inelastic scattering is one of the frontiers 
in perturbative QCD (pQCD).  This corresponds to the  small Bjorken $x$ region
and  
represents the challenge of studying the interface between perturbative and 
non-perturbative QCD, with the peculiar feature that this transition is 
taken in a kinematical region where the strong coupling constant $\alpha_s$ is small.
The transition between the perturbative regime and the non-perturbative regime
may provide an understanding of the soft interactions in the language of 
QCD and establish whether the Regge model can be justified from QCD 
principles.
From the 
experimental side, data from electron-proton collider HERA at DESY ($\sqrt{s}=300\,GeV$), provide new insights into the structure of the proton. The high energy allows to probe the proton in hitherto unexplored kinematical regions. The collider   HERA allows the measure of  structure functions 
$F_2,\,F_2^c$, $F_L$ and the cross sections of  
vector meson production and diffractive processes  for Bjorken $x$ values down to $10^{-4}$. 

In the region of moderate $x$ ($x\ge 10^{-2}$) the well-established methods 
of operator expansion and renormalization group equations have been applied 
successfully. The DGLAP equations \cite{dglap}, which are based upon the sum of QCD 
ladder diagrams, are the evolution equations in 
this kinematical region. However, in the small $x$ region, the density of 
gluons and quarks becomes very high and a new dynamical effect is expected 
to stop the further growth of the structure functions.  
In this kinematical region we are dealing with a system of partons which are 
still at small distances, where  $\alpha_s$ is still 
small, but the density of partons becomes so large that the usual methods of 
pQCD cannot be applied \cite{forever}. About seventeen years ago, 
Gribov {\it et al.} \cite{100} have performed a detailed study of this region. 
They  argued that  the physical processes of interaction and recombination of 
partons become important in the parton cascade at a large value of the parton 
density, and that these shadowing corrections could be expressed in a new evolution equation - the GLR equation. This equation considers the leading non-ladder contributions:
the multi-ladder diagrams, denoted as fan diagrams. The main characteristics of this 
equation are:  it predicts a saturation of the gluon distribution at very small 
$x$;  it predicts a critical line, separating the perturbative regime from the 
saturation regime; it is only valid in the border of this critical line. Therefore,
the GLR equation predicts the limit of its validity.
In the last decade,  the solution \cite{collins,bartels1,bartels2} and 
possible generalizations  \cite{bartels,laenen} of the GLR equation have 
been studied in great detail. Recently, an eikonal approach to  the  
shadowing corrections was proposed in the literature \cite{ayala1,ayala2,ayala3}. 
 The starting point of these papers is the proof of the Glauber formula in QCD \cite{mueller}, which considers only the interaction  of the fastest partons  with the target. As in QCD we rather expect that all partons should interact with the target, a more general approach was proposed in \cite{ayala1}. This approach was applied to the nuclear case in \cite{ayala1} and to the 
nucleon case in \cite{ayala2,ayala3}. Some of the  main characteristics of this 
approach are  its validity in a large kinematic region and that it provides  a limit case of the GLR equation. 
In this paper the shadowing corrections will be estimated using the eikonal 
approach. In the next section we discuss this approach in more detail,  
pointing first our motivation.

 One of the most striking discoveries at  HERA is the steep rise 
of the proton structure function $F_2(x,Q^2)$ with decreasing Bjorken $x$ 
\cite{h1}.
The behaviour of the structure function at small $x$ is driven by the gluon 
through the process $g\rightarrow q\overline{q}$.
Therefore the gluon distribution is the observable that governs the physics of high energy processes in QCD.
 HERA  shows that the deep inelastic structure function $F_2(x,Q^2)$
has a steep behaviour in the small-$x$ region $(10^{-2} > x > 10^{-5})$, even
for  very small virtualities  $(Q^2 \approx 1\,GeV^2)$. Indeed, considering
$F_2 \propto x^{- \lambda}$ at low $x$, the HERA data are consistent with a 
$\lambda$ that varies from 0.15 at $Q^2 = 0.85\,GeV^2$ to 0.4 at $Q^2 = 20 \,GeV^2$.
This steep behaviour is well described in the framework of the DGLAP evolution
equations  with an appropriate choice of input distributions and the choice of 
the starting scale for the $Q^2$ evolution, by all groups doing  global 
fits of the data \cite{grv95,mrs,cteq}. No other ingredient is needed to describe 
the experimental data.
Using this result is it possible to conclude that there will be no significant 
shadowing contributions at  HERA small-$x$ regime?
Here, we address this question. Clearly, the answer of this question is 
connected with the demonstration that the corrections due to
shadowing corrections (SC) are negligible at least at  HERA kinematic region.
If it is not so the DGLAP approach is not better or worse than any other 
evolution mechanism  developed to describe the experimental data. 

The SC for the $F_2(x,Q^2)$ 
proton structure function and $xG(x,Q^2)$  gluon distribution were computed 
considering the eikonal approach  in refs. \cite{ayala2,ayala3} .
In those works, the authors estimate the value of the SC which turn out to be 
essential  in the gluon distribution but rather small in $F_2$. 
Consequently, the $F_2$ data cannot determine if there is a new dynamical effect at 
HERA. Therefore, to obtain a more precise evidence of SC 
at HERA kinematic region, we must consider other observables directly dependent 
on the behaviour of the gluon distribution, and consequently, more sensitive to SC. 
Some of these observables are
$F_L$, $F_2^c$, $J/\Psi$ production,  diffractive leptoproduction  of vector
mesons and open charm production.
 In this paper we consider the observables 
 $F_L$ and $F_2^c$. 
The dominant  charm production  mechanism is the boson-gluon fusion. 
Therefore, the charm component $F_2^c$ of the structure function is a 
sensitive probe of the gluon distribution at small $x$. Furthermore, the 
violation of the Callan-Gross relation ($F_L=F_2 - 2xF_1=0$) occurs when 
the quarks acquire transverse momenta from QCD radiation. Consequently, also 
the longitudinal structure function  is a sensitive probe of the gluon 
distribution at small $x$.
In this work we estimate the SC in  $F_L$ and $F_2^c$  in the approach 
proposed in \cite{ayala2}. We calculate these observables using the Altarelli-Martinelli equation \cite{alta} and the boson-gluon fusion cross section \cite{grv95}, respectively. The gluon distribution and the $F_2$ structure function necessary to these equations are calculated considering the eikonal approach.
Actually, there are few data in these observables, but more measurements, with
better statistics, will be available in the next years. Therefore it is very 
important the analysis of the  behaviour of these observables considering SC.

This paper is organized as follows. In Section II the eikonal approach and the shadowing corrections for the $F_2(x,Q^2)$ and $xG(x,Q^2)$ 
are briefly presented. In  Section III we obtain the formulae for the calculation of  
the $F_L$ and $F_2^c$ and we apply our results to  HERA 
data. Moreover, we compare our results  with the  predictions of DGLAP dynamics. Finally, in Section IV, we present 
our  conclusions.

\section{The Eikonal Approach in pQCD}

The space-time picture of the Eikonal approach in the target
rest frame can be viewed as the decay of the virtual gluon at high energy
(small $x$) into a gluon-gluon pair long before the 
interaction with the target. The $gg$ pair subsequently interacts 
with the target.  In the small $x$ region, where 
$x << \frac{1}{2mR}$ ($R$ is the size of the target), the $gg$ pair 
crosses the target with fixed
transverse distance $r_t$ between the gluons. 

The cross section of the absorption of a gluon $g^*$ with virtuality $Q^2$ can be written as 
\begin{eqnarray}
\sigma^{g^* + nucleon}(x,Q^2) = \int_0^1 dz \int \frac{d^2r_t}{\pi} 
\int \frac{d^2b_t}{\pi} |\Psi_t^{g^*}(Q^2,r_t,x,z)|^2 \sigma^{gg+nucleon}(z,r_t^2)\,\,,
\label{sec1}
\end{eqnarray}
where $z$ is the fraction of energy carried by the gluon, $b_t$ is the impact parameter and $\Psi_t^{g^*}$ is the wave function of the transverse polarized gluon in the virtual probe. Furthermore, $\sigma^{gg+nucleon}(z,r_t^2)$ is the cross section of the interaction of the $gg$ pair with the  nucleon. In the leading log$(1/x)$ approximation we can neglect the change of $z$ during the interaction and describe the cross section $\sigma^{gg+nucleon}(z,r_t^2)$ as a function of the variable $x$.

Using the unitarity in the $s$-channel, the  cross section $\sigma^{gg+nucleon}(z,r_t^2) = \sigma(x,r^2_t)$ can be written in the form
\begin{eqnarray}
\sigma(x,r^2_t) = 2 \int d^2b_t \, (1 - e^{-\frac{1}{2} \Omega (x,r_t,b_t)})\,\,,
\label{sig}
\end{eqnarray}
where $\Omega$ is an arbitrary real function, which can be specified
only in a more detailed theory or approach than the unitarity constraint. One 
of such specified model is the Eikonal model.

The Eikonal model assumes that $\Omega$ is small $(\Omega << 1)$ and its 
$b_t$ dependence can be factorized as $\Omega = \overline{\Omega} S(b_t)$, with
the normalization $\int d^2b_t\, S(b_t) = 1$. The factorization was proven for
the DGLAP evolution equations \cite{100} and, therefore, all our further 
calculations will be valid for the DGLAP evolution equations in the region of small
$x$ or, in other words, in the double log approximation (DLA) of perturbative
QCD (pQCD). The Eikonal approach is the assumption that 
$\Omega = \overline{\Omega} S(b_t)$ in the whole kinematical region. 


In \cite{plb} the authors demonstrate that $\overline{\Omega}$ is given by
\begin{eqnarray}
 \overline{\Omega} = \sigma_N^{gg} = \frac{3 \alpha_s(\frac{4}{r_t^2})}{4}\,\pi^2\,r_t^2\,
 xG(x,\frac{4}{r_t^2})\,\,,
\label{omega}
\end{eqnarray}
where  $xG(x,Q^2)$ is the gluon distribution.  Therefore the behaviour of the cross section (\ref{sig}) in the small-$x$ region is determinated by the behaviour of the gluon distribution in this region.

Considering the $s$-channel unitarity and the eikonal model, equation  (\ref{sec1}) can be written as
\begin{eqnarray}
\sigma^{g^* + nucleon}(x,Q^2) = \int_0^1 dz \int \frac{d^2r_t}{\pi} 
\int \frac{d^2b_t}{\pi} |\Psi_t^{g^*}(Q^2,r_t,x,z)|^2 
\,(1 - e^{-\frac{1}{2}  \overline{\Omega} S(b_t)})\,\,,
\label{diseik}
\end{eqnarray}
Using the relation $\sigma^{g^* + nucleon}(x,Q^2) = \frac{4\pi^2 \alpha_s}{Q^2}xG(x,Q^2)$ and the expression of the wave $\Psi^{g^*}$ calculated in \cite{ayala1},  the Glauber-Mueller formula for the gluon  distribution is obtained as
\begin{eqnarray}
xG(x,Q^2) = \frac{4}{\pi^2} \int_x^1 \frac{dx^{\prime}}{x^{\prime}}
\int_{\frac{4}{Q^2}}^{\infty} \frac{d^2r_t}{\pi r_t^4} \int_0^{\infty}
\frac{d^2b_t}{\pi}\,2\,[1 - e^{-\frac{1}{2}\sigma_N^{gg}(x^{\prime}
,\frac{r_t^2}{4})S(b_t)}]\,\,.
\label{gluon}
\end{eqnarray}

The use of the Gaussian parameterization for
the  nucleon profile function $S(b_t) = \frac{1}{\pi R^2} e^{-\frac{b^2}{R^2}}$ simplifies the calculations. Consequently, doing the integral over $b_t$,  the  master equation is obtained
\begin{eqnarray}
xG(x,Q^2) = \frac{2R^2}{\pi^2}\int_x^1 \frac{dx^{\prime}}{x^{\prime}}
\int_{\frac{1}{Q^2}}^{\frac{1}{Q_0^2}} \frac{d^2r_t}{\pi r_t^4} \{C + ln(\kappa_G(x^{\prime}, r_t^2)) + E_1(\kappa_G(x^{\prime}, r_t^2))\}  \,\,,
\label{master}
\end{eqnarray} 
where $C$ is the Euler constant,  $E_1$ is the exponential function,  and the function  $\kappa_G(x, r_t^2) = \frac{3 \alpha_s}{2R_A^2}\,\pi\,r_t^2\,
 xG(x,\frac{1}{r_t^2})$. If equation (\ref{master}) is expanded for small $\kappa_G$, the first term (Born term) will correspond to the usual DGLAP equation in the small $x$ region, while the other terms will take into account the shadowing corrections.

The master formula  (\ref{master}) is correct in the double logarithmic approximation (DLA) \cite{ayala2}. As shown in \cite{ayala2} the DLA does not work quite well in the accessible kinematic region ($Q^2 > 1\,GeV^2$ and $x > 10^{-4}$). Consequently, a more realistic approach must be considered to calculate the gluon distribution $xG$. In \cite{ayala2} the subtraction of the Born term of   (\ref{master}) and addition of the GRV parameterization was proposed. This procedure gives
\begin{eqnarray}
xG(x,Q^2) & = & xG^{master}  [\mbox{eq. (\ref{master})}] + xG^{GRV}(x,Q^2) \nonumber \\
&  - & \frac{\alpha_s N_c}{\pi} 
\int_x^1 \int_{Q_0^2}^{Q^2}  \frac{dx^{\prime}}{x^{\prime}}  \frac{dQ^{\prime 2}}{Q^{\prime 2}}\,x^{\prime}G^{GRV}(x^{\prime},Q^{\prime 2})\,\,.
\label{gluon2}
\end{eqnarray}
The above equation includes also $xG^{GRV}(x,Q_0^2)$ as the initial condition for the gluon distribution and gives  $xG^{GRV}(x,Q^2)$ as the first term of the expansion with respect to $\kappa_G$. Therefore, this equation is an attempt to include the full expression for the anomalous dimension for the scattering off each nucleon, while the use of  the DLA  takes into account all SC. In \cite{ayala1} this procedure was applied to obtain the shadowing corrections to the nuclear gluon distribution and in \cite{ayala2} to the nucleon gluon distribution.

The equation (\ref{gluon}) is not a non-linear equation type the GLR equation \cite{100}; it is the analogue of the Glauber formula, which provides the possibility to  calculate the SC using the solution of the DGLAP evolution equation. In this paper, as shown in eq. (\ref{gluon2}), we use the GRV parameterization as a solution of the DGLAP evolution equation. It  describes all available experimental data quite well \cite{h1}. It should be also stressed here that we disregard how much of SC has been taken into account in this parameterization in the form of the initials distributions.

A similar
approach can be used to obtain the SC
 for the deep inelastic structure function $F_2(x,Q^2)$. This was made in the references \cite{ayala2,plb}.
The main result of these references is that the structure function $F_2$ can be written, in the eikonal approach, as
\begin{eqnarray}
F_2(x,Q^2) =  \frac{N_c}{6\pi^3} \sum_{u,d,s} \epsilon_q^2 \int_{\frac{1}{Q^2}}^{\infty} \frac{d^2r_t}{r_t^4} \int d^2b_t 
\{1 - e^{-\frac{1}{2}\Omega_{q\overline{q}}(x,r_t,b_t)}\}\,\,,
\label{f2eik}
\end{eqnarray}
where $\Omega_{q\overline{q}}=\frac{4}{9}\overline{\Omega}S(b_t)$ (see eq. (\ref{omega})). 

 Following the same steps used in the case of the  gluon distribution,  the proton structure function  can be written as
\begin{eqnarray}
F_2(x,Q^2) =  \frac{2R^2}{3\pi^2} \sum_{u,d,s} \epsilon_q^2 \int_{\frac{1}{Q^2}}^{\frac{1}{Q_0^2}} \frac{d^2r_t}{\pi r_t^4} \{C + ln(\kappa_q(x, r_t^2)) + E_1(\kappa_q(x, r_t^2))\}\,\,,
\label{diseik2}
\end{eqnarray}
where $\kappa_q = \frac{4}{9} \kappa_G$.
Similarly as made in the case of the  gluon distribution, to obtain a more realistic approach the Born term must be subtracted and the GRV parameterization must be summed. Therefore the proton structure function is given by
\begin{eqnarray}
F_2(x,Q^2) = F_2(x,Q^2) [\mbox{eq. (\ref{diseik2})}] - F_2(x,Q^2)[\mbox{Born}] + F_2(x,Q^2) [\mbox{GRV}] \,\,,
\label{f2ab}
\end{eqnarray}
where $ F_2(x,Q^2)[\mbox{Born}]$ is the first term in the expansion in $\kappa_q$ of the equation (\ref{diseik2}), and 
\begin{eqnarray}
F_2(x,Q^2) [\mbox{GRV}] = \sum_{u,d,s} \epsilon_q^2 \,[xq(x,Q^2) + x\overline{q}(x,Q^2)] + F_2^c(x,Q^2)
\end{eqnarray}
is calculated using the GRV parameterization.

In  \cite{ayala2} the shadowing corrections for the $F_2(x,Q^2)$ 
proton structure function and $xG(x,Q^2)$  gluon distribution were computed 
considering the eikonal approach, and
 the SC which turn out to be 
essential  in the gluon distribution are rather small in $F_2$.

\begin{figure}
\centerline{\psfig{file=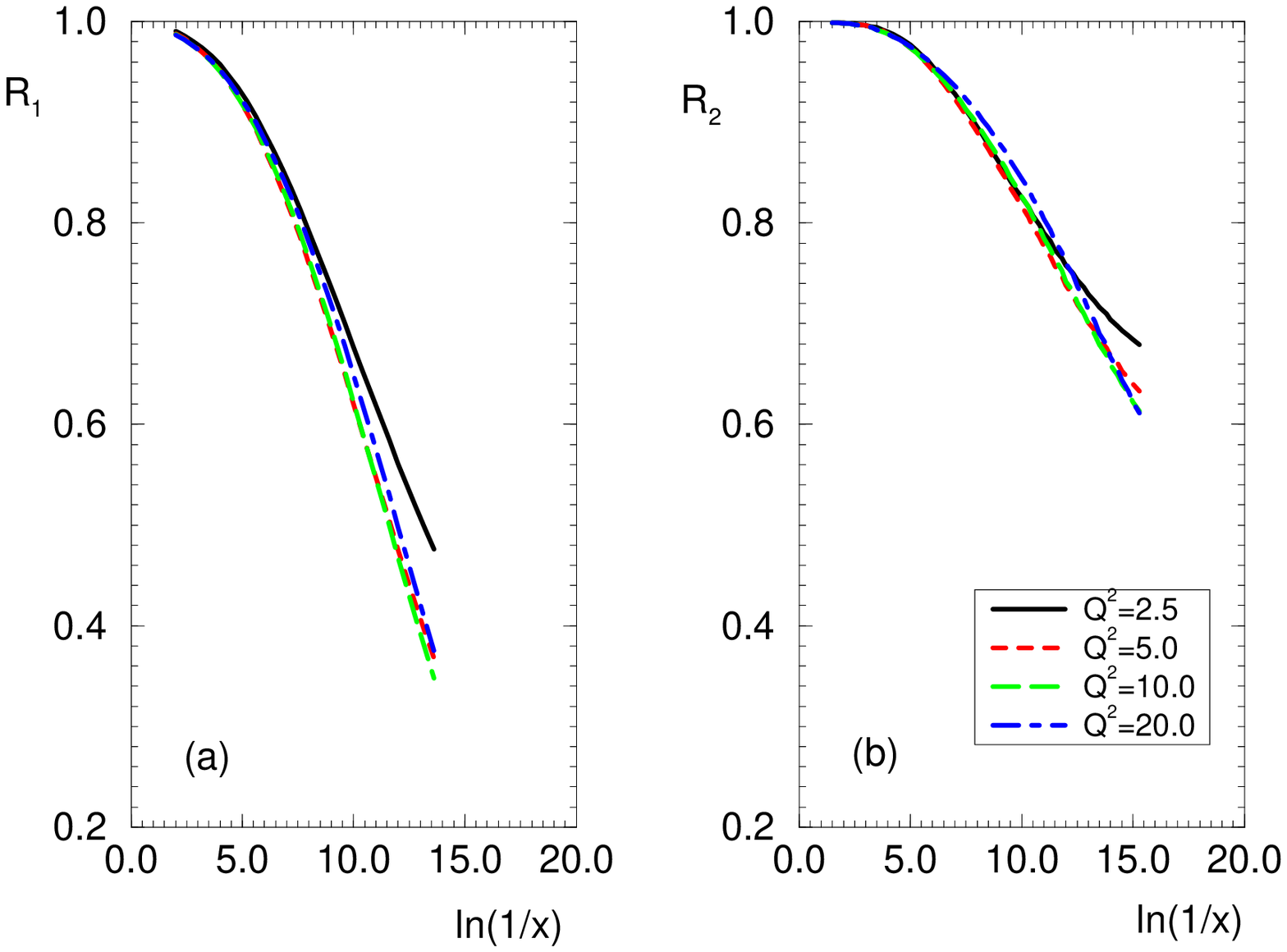,width=150mm}}
\caption{ (a) The ratio $R_1 = \frac{xG^{GM}}{xG^{GRV}}$  and (b) the ratio $R_2 = \frac{F_2^{GM}}{F_2^{GRV}}$  as a function of the variable $ln(\frac{1}{x})$ for different virtualities.}
\label{fig1}
\end{figure}

The  result above is demonstrated in  figure
\ref{fig1}, where  we present the ratios
\begin{eqnarray}
R_1=\frac{xG^{GM}(x,Q^2)}{xG^{GRV}(x,Q^2)}
\label{r1}
\end{eqnarray}
and  
\begin{eqnarray}
R_2=\frac{F_2^{GM}(x,Q^2)}{F_2^{GRV}(x,Q^2)}\,\,,
\label{r2}
\end{eqnarray}
as a function of variable $ln(\frac{1}{x})$ for different values of $Q^2$.
In this case $GM$ represents that  the function ($F_2$ or $xG$) was obtained 
using the eikonal approach, and $GRV$ represents that  the function  was obtained using the GRV parameterizations \cite{grv95}. We can see that in HERA kinematical region $(3 \le \,ln\,\frac{1}{x}\, \le 12)$ the behaviour of the gluon distribution is strongly modified by shadowing corrections. In \cite{ayala2} the authors demonstrated that the $F_2$ data can be described considering shadowing corrections, {\it i.e.} the $F_2$ data cannot determinate if there is a new dynamical effect at HERA. Consequently, observables
directly dependent on the behaviour of the gluon distribution must be considered to 
discriminate the SC.
From results obtained in \cite{ayala2} we conclude that the eikonal model provides a good description of the SC and can be taken as a correct approximation in the approach to the nucleon case for  HERA kinematical region.
In the next section we estimate the shadowing corrections to observables directly dependent on the behaviour of the gluon distribution.

In this paper we estimate the shadowing corrections using an eikonal 
approach proposed in \cite{ayala2}.  The eikonal approach gives  
sufficiently reliable results for the HERA kinematic region, however, 
it is not the most efficient way to calculate the shadowing corrections. 
In \cite{ayala2}, a generalized equation which takes into account the interaction 
of all partons in a parton cascade with the target was proposed. 
The main properties of generalized equation are: (i) the iterations of this 
equation coincide with the iteration of the Glauber-Mueller formula; 
(ii) its solution matches  the solution of the DGLAP evolution equation 
in the double-logarithmic-approximation (DLA) limit of pQCD; (iii) it 
has the GLR equation as a  limit, and (iv)  contains the 
Glauber-Mueller formula. Therefore, the generalized equation is valid in a 
large kinematic region. In this paper we have used the Glauber-Mueller 
formula, since that in the HERA kinematic region the solutions of the 
generalized equation and of Glauber-Mueller formula are approximately 
identical \cite{ayala2}.   However, a more accurate approach, to a  larger kinematic 
region than the HERA one, should consider as an input the solution 
of the generalized equation, in the future \cite{talklev}.

\section{$F_L$ and $F_2^c$  }

Our goal in this section is the determination of small-$x$ shadowing 
corrections at $F_L$ and $F_2^c$  using the approach proposed in 
\cite{ayala2}. These observables are directly dependent on the behaviour of the gluon 
distribution, and consequently, more sensitive to SC. 

Recently, the SC to the $F_2$ slope was estimated \cite{lgm2}. This observable is also  directly dependent on the behaviour of the  gluon distribution \cite{victor}. It was shown that the contribution of SC is large in the $F_2$ slope and that the experimental data can be described incorporating the SC. This result is a strong motivation to estimate the SC at $F_L$ and $F_2^c$.

\subsection{The longitudinal structure function}

The longitudinal structure function in deep inelastic scattering is one 
of the observables from which the gluon distribution can be unfolded. Its 
experimental determination is difficult since it usually requires cross 
sections measurements at different values of  center of mass energy, 
implying a change of beam energies. A direct change of  the beam energies 
at HERA has been widely discussed \cite{hera96}. An alternative possibility 
is to apply the radiation of a hard photon by the incoming electron. Such hard 
radiation results into an effective reduction of the center of mass energy. Several 
studies on the use of  such events to measure $F_L$ have  been carried out 
\cite{favart}. 
With  these measurements, which in principle could be performed in the near future, it may be possible to explore the structure of $F_L(x,Q^2)$ in the low $x$ range.

Current measurements of $R(x,Q^2) = F_L/(F_2 - F_L)$ have been made by 
various fixed target lepton-hadron scattering experiments at higher $x$ 
values \cite{arneodo}. At low $x$, the measurements of this observable are 
very scarce. Recently, the H1 Collaboration has published the first 
$F_L$ data at small $x$. 
To obtain the data, the  structure function $F_2$ was parameterized taken 
only  data for $y < 0.35$, where the  contribution of $F_L$ is small. The quantity 
$y=\frac{Q^2}{sx}$ describes in the rest frame  
of the proton, the energy transfer from the incoming to the outgoing electron. 
This parameterization was evolved in $Q^2$ according to the DGLAP evolution 
equations. This provides predictions for the structure function $F_2$ in the 
high $y$ region which allowed, by subtraction of the contribution of $F_2$ 
to the cross section, the determination of the longitudinal structure function.
Therefore, in order to constrain the value of $F_L$ in low $x$ 
regime explored by HERA, the H1 Collaboration extracted $F_L$ assuming  that 
the  DGLAP evolution holds. 
Our point of view agrees with Thorne's statement \cite{thorne}, pointing that the 
assumption of the correctness of the DGLAP evolution equations at small $x$ 
implies that $F_L$ is already isolated. Therefore, our comparisons with these data
must be considered as an estimate of the SC compared with the results of the 
standard evolution equations.

One can write the longitudinal structure function $F_L$ in terms of the cross section for the absorption of longitudinally polarized photons as
\begin{eqnarray}
F_L(x,Q^2)  =  \frac{Q^2(1-x)}{4\pi^2 \alpha_{em}}\,\sigma_L(x,Q^2)
\approx \frac{Q^2}{4\pi^2 \alpha_{em}}\,\sigma_L(x,Q^2)
\label{fl}
\end{eqnarray}
at small $x$.
Longitudinal photons have zero helicity  and can exist only virtually. In the Quark-Parton Model (QPM), helicity conservation of the electromagnetic vertex yields the Callan-Gross relation, $F_L=0$, for scattering on quarks with spin $1/2$. This does not hold when the quarks acquire transverse momenta from QCD radiation. Instead, QCD yields the Altarelli-Martinelli equation\cite{alta}
\begin{eqnarray}
F_L(x,Q^2) = \frac{\alpha_s(Q^2)}{2\pi}\,x^2\, \int_x^1 \frac{dy}{y^3}[\frac{8}{3}\,F_2(y,Q^2) + 4\,\sum_q e_q^2 (1-\frac{x}{y})yg(y,Q^2)]\,\,,
\label{flalta}
\end{eqnarray}
expliciting the dependence of $F_L$ on the strong constant coupling and the gluon density. At small $x$ the second term with the gluon distribution is the dominant one. Consequently, the expression (\ref{flalta}) can be approximated reasonably by  $F_L(x,Q^2) \approx 0.3\, \frac{4 \alpha_s}{3 \pi} xg(2.5x,Q^2)$ \cite{cooper}. This equation demonstrates  the close relation between the longitudinal structure function and the gluon distribution. Therefore, we expect that the longitudinal structure function to be sensitive to the shadowing corrections at HERA kinematic region. In this paper we  calculate $F_L$ using the Altarelli-Martinelli equation  (\ref{flalta}).

\begin{figure}
\centerline{\psfig{file=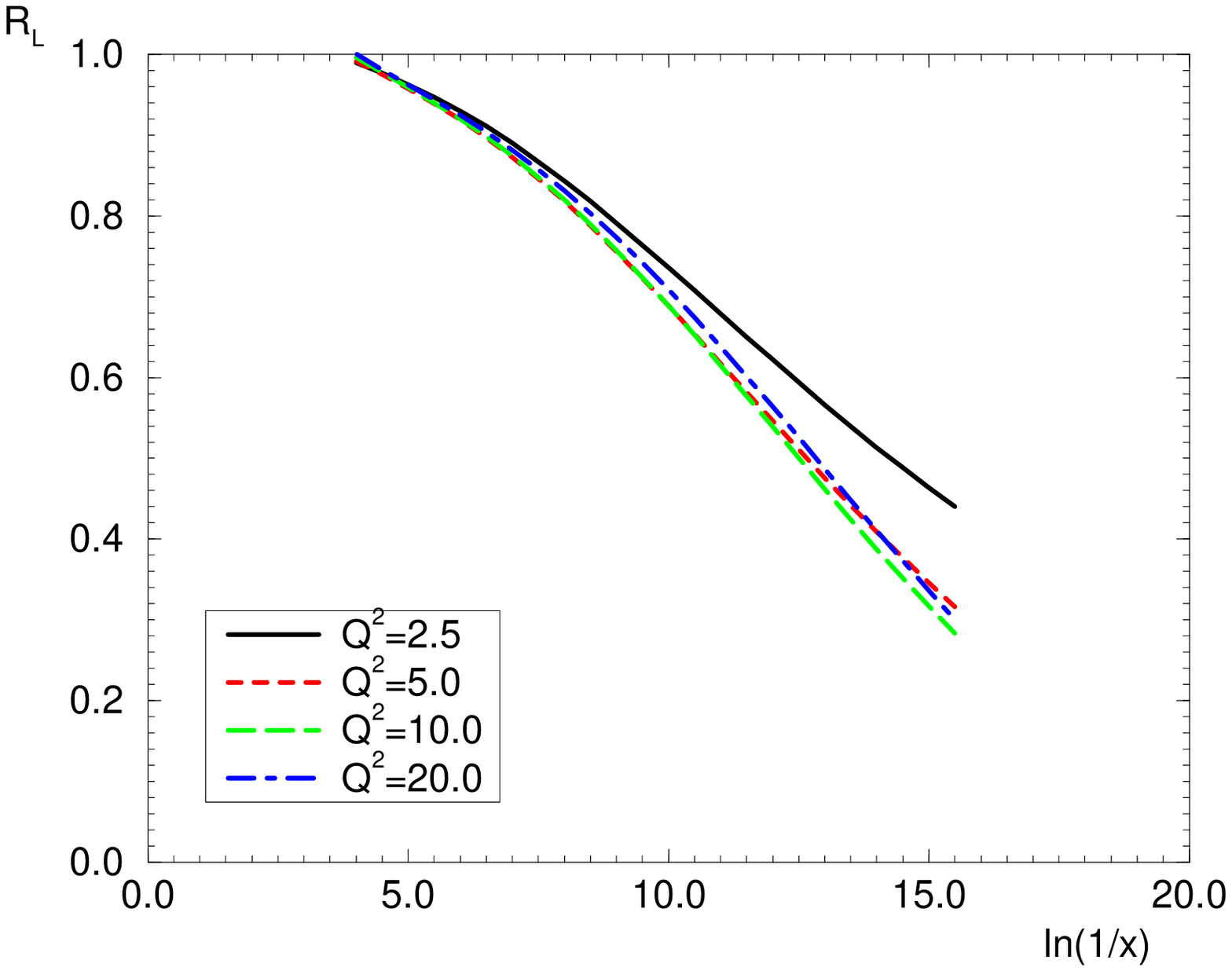,width=150mm}}
\caption{ The ratio $R_L = \frac{F_L^{GM}}{F_L^{GRV}}$  as a function of the variable $ln(\frac{1}{x})$ for different virtualities. }
\label{fig2}
\end{figure}

Considering the expressions  (\ref{gluon2}), (\ref{f2ab}) and (\ref{flalta}) we can estimate the SC for the $F_L$ structure function. In  figure \ref{fig2} we present the ratio
\begin{eqnarray}
R_L= \frac{F_L^{GM}(x,Q^2)}{F_L^{GRV}(x,Q^2)}\,\,,
\label{rl}
\end{eqnarray}
where $F_L^{GM}(x,Q^2)$ indicates that the longitudinal structure function was obtained using the gluon distribution solution of the Glauber-Mueller formula, eq. (\ref{gluon}), and $F_L^{GRV}(x,Q^2)$ indicates that $F_L$ was obtained using the GRV parameterization.
We can see that the behaviour of  $F_L$ is strongly modified by shadowing corrections. The suppression increases with
$ln(\frac{1}{x})$ and is much bigger than for the $F_2$ case. 
In the region of  HERA data, $3 \le\,ln(\frac{1}{x})\le\,12$, the shadowing corrections are not bigger than $44\%$ ($ln(\frac{1}{x}) \approx 12$). The SC are as  big as  $70\%$ only at very small value of $x$ ($ln(\frac{1}{x}) \approx 15$), where we have no experimental data.

\begin{figure}
\centerline{\psfig{file=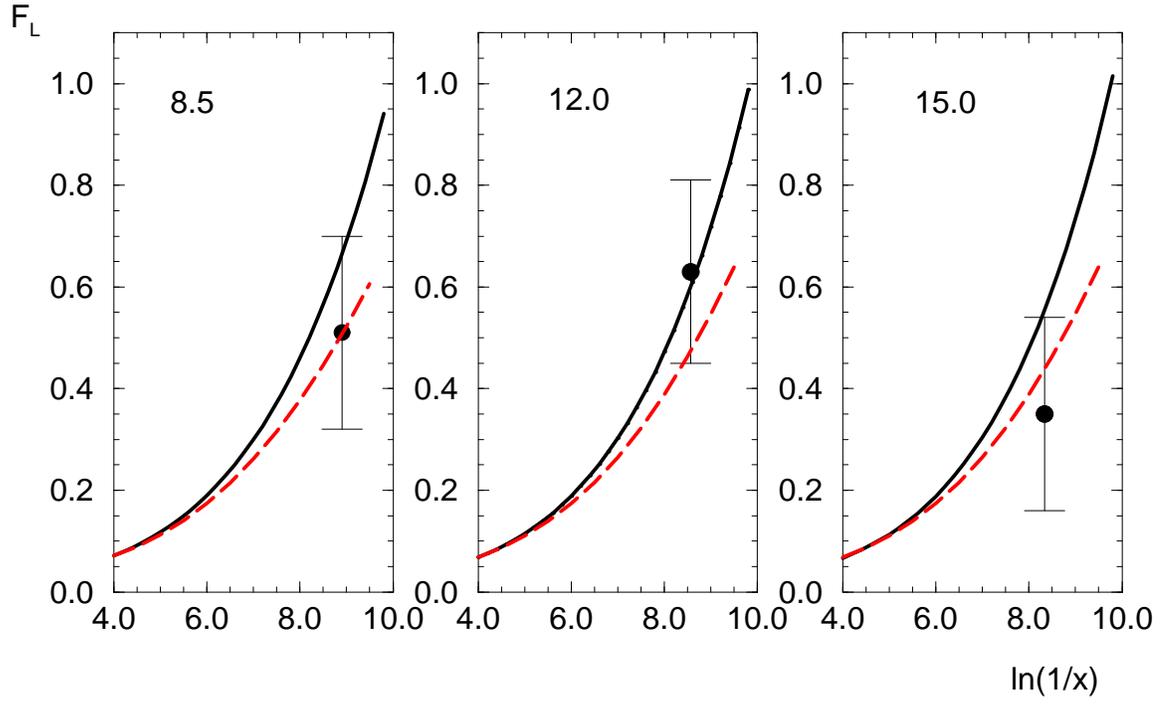,width=150mm}}
\caption{ The longitudinal structure function $F_L$ for $Q^2=8.5,\,12,\,15\,GeV^2$. The solid curve represents that the $F_L$ was calculated using the GRV parameterization and the dashed curve that its was calculated using $xG$ and $F_2$ obtained in the eikonal model. Data from H1.}
\label{fig3}
\end{figure}

\begin{figure}
\centerline{\psfig{file=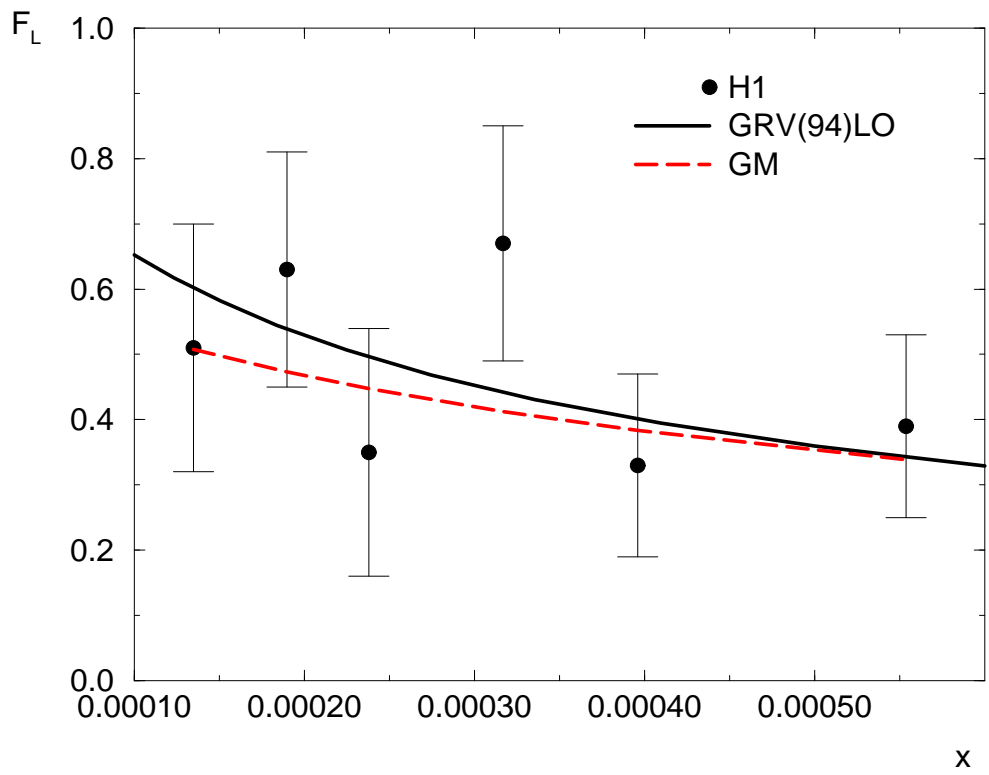,width=150mm}}
\caption{ The longitudinal structure function $F_L$ for all H1 data. }
\label{fig4}
\end{figure}

In  figure \ref{fig3} we present the behaviour of the $F_L$ structure function 
with (dashed curve) and without shadowing (solid curve) as a function of 
$ln(\frac{1}{x})$ for different virtualities. We compare our results with 
the scarce H1 data \cite{flh1}. We can see that both curves agree with the data, however the experimental error are still very large. 
In  figure \ref{fig4}  we present our results for $F_L$ and compare with all H1 data. From the above results we can conclude that the eikonal model gives a good description of the longitudinal structure function and describes the experimental data. However, new data, with better statistics, could verify the shadowing corrections at HERA kinematic region.

Our conclusion of this section is that the  longitudinal structure function 
$F_L$ is a good observable to isolate the shadowing corrections at HERA. 
Consequently, we must stress the importance of measuring directly $F_L$ at HERA
as a probe of the dynamics of small $x$ physics.
We  hope this result could also  motivate the acquisition of new data in the 
next years.

\subsection{ The charm 
component $F_2^c$ of the structure function}

The problem of how to treat heavy quark contributions in the deep inelastic 
structure functions has been widely  discussed, see for example \cite{neerven}.
It has been brought into focus recently by the very precise $F_2(x,Q^2)$ 
data from HERA \cite{h1}. Both the H1 and ZEUS collaborations have measured the charm 
component $F_2^c$ of the structure function at small $x$ and have found it 
to be a large (approximately $25\%$) fraction of the total. This is in sharp 
contrast to what is found at large $x$, where typically $F_2^c/F_2\, \approx 
{\cal{O}}(10^{-2})$.

For the treatment of the charm component of the structure function there are 
basically two different prescriptions for the charm production in the literature.
The first one is advocated in \cite{grvc} where the charm quark is treated 
as a heavy quark and its contribution is given by fixed-order perturbation 
theory. This involves the computation of the boson-gluon fusion process. In the 
other approach \cite{acot} the charm is treated  similarly  to a massless
quark and its contribution is described by  a parton density in a hadron.
Here our goal is to obtain the shadowing corrections to the charm 
structure function in the eikonal approach, without considering in detail the question of how to treat heavy quark contributions  \cite{neerven}.
However, some comments are important.  We consider the charm production via 
boson-gluon fusion, where the charm is treated as a heavy quark and not a 
parton. This scheme is usually  called  Fixed Flavour Number Scheme (FFNS).  
In this scheme, by definition, only light partons (e.g. $u$, $d$, $s$ and $g$)  are included in the initial state for
charm production: the number of parton flavours
$n_f$ is kept at a fixed value regardless  the energy scales  involved.
The boson-gluon fusion gives the correct description of $F_2^c$ for $Q^2 < 4m_c^2$ and should remain a reasonable approximation  to $F_2^c$ for $Q^2 \geq 4m_c^2$. However, the boson-gluon fusion model will inevitably break down at larger $Q^2$ values as the charm can no longer be treated as a non-partonic heavy object, and  begins to evolve more like the lighter components of the quark sea. Therefore, our estimates in this  scheme should be considered with caution in  the region of large $Q^2$.

A $c\overline{c}$ pair can be created  by boson-gluon fusion when  the squared  invariant mass  of the hadronic final state $W^2 \ge 4m_c^2$. Since 
$W^2 = \frac{Q^2(1-x)}{x} + M_N^2$, where $M_N$ is the nucleon mass, the charm production  can  occur well below the $Q^2$ threshold, $Q^2 \approx  4m_c^2$, at small $x$. 
The charm contribution to the proton structure function, in leading order (LO), is given by \cite{grv95}
\begin{eqnarray}
\frac{1}{x} F_2^c(x,Q^2,m_c^2) = 2 e_c^2 \frac{\alpha_s(\mu^{\prime 2})}{2\pi} 
\int_{ax}^1 \frac{dy}{y}\, C_{g,2}^c(\frac{x}{y},\frac{m_c^2}{Q^2})\,g(y,\mu^{\prime 2})  \,\,,
\label{f2c}
\end{eqnarray}
where $a=1+\frac{4m_c^2}{Q^2}$ and the factorization scale $\mu^{\prime}$ is 
assumed $\mu^{\prime 2}=4m_c^2$.  $C_{g,2}^c$ is the coefficient function 
given by
\begin{eqnarray}
C_{g,2}^c(z, \frac{m_c^2}{Q^2})  & = & \frac{1}{2} \{ [z^2 + (1-z)^2 +z(1-3z)\frac{4m_c^2}{Q^2} - z^2 \frac{8m_c^4}{Q^4}]
ln \frac{1+\beta}{1-\beta} \nonumber \\ & + & \beta[-1 +8z(1-z) -z(1-z)\frac{4m_c^2}{Q^2}]\}\,\,,
\end{eqnarray}
where $\beta= 1 - \frac{4m_c^2 z}{Q^2 (1-z)}$  is the velocity of one of the charm quarks in the boson-gluon center-of-mass 
frame.
Therefore, in leading order, ${\cal{O}}(\alpha_s)$, $F_2^c$ is directly sensitive only to the gluon density via the well-known Bethe-Heitler process
$\gamma^*g \rightarrow c\overline{c}$.
The dominant uncertainty in the QCD calculations arises from the uncertainty 
in the  charm quark mass. In this paper we assume $m_c = 1.5\,GeV$.

\begin{figure}
\centerline{\psfig{file=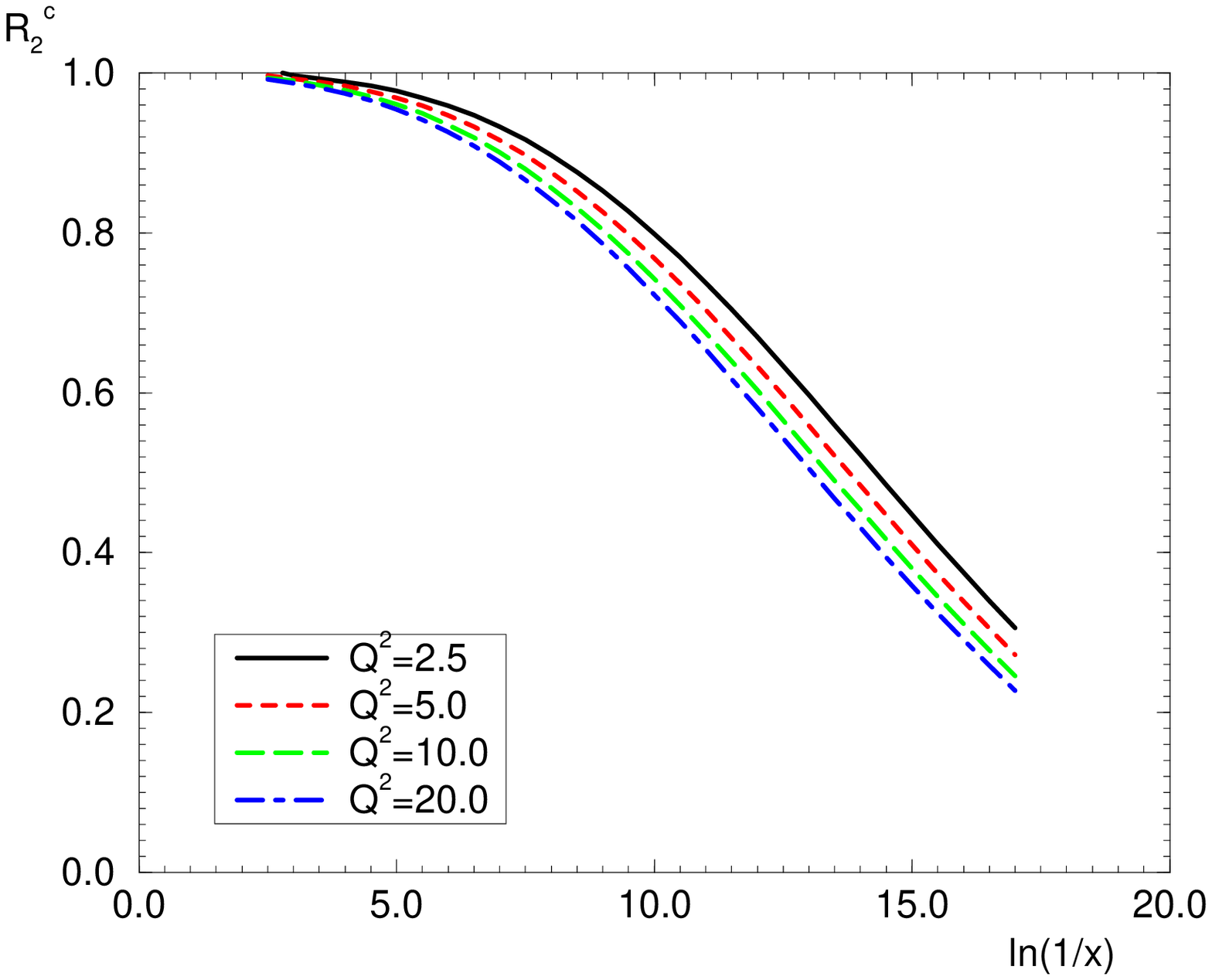,width=150mm}}
\caption{The ratio $R_2^c = \frac{F_2^{c\,GM}}{F_2^{\c,GRV}}$  as a function of the variable $ln(\frac{1}{x})$ for different virtualities.}
\label{fig5}
\end{figure}

Considering the expressions  (\ref{gluon2}) and (\ref{f2c}) we can estimate the shadowing corrections for the $F_2^c$ structure function. In  figure \ref{fig5} we present the ratio
\begin{eqnarray}
R_2^c= \frac{F_2^{c\,GM}(x,Q^2)}{F_2^{c\,GRV}(x,Q^2)}\,\,.
\label{r2c}
\end{eqnarray}
We can see that the behaviour of the $F_2^c$ is strongly modified by the SC.  The suppression due to shadowing corrections increases with
$ln(\frac{1}{x})$ and is much bigger than for the $F_2$ case. In the region of  HERA data, $3 \le\,ln(\frac{1}{x})\le\,12$, the SC are not bigger than $40\%$ ($ln(\frac{1}{x}) \approx 12$). The SC are bigger than $62\%$ only at very small value of $x$ ($ln(\frac{1}{x}) \approx 15$), where we have no experimental data.

\begin{figure}
\centerline{\psfig{file=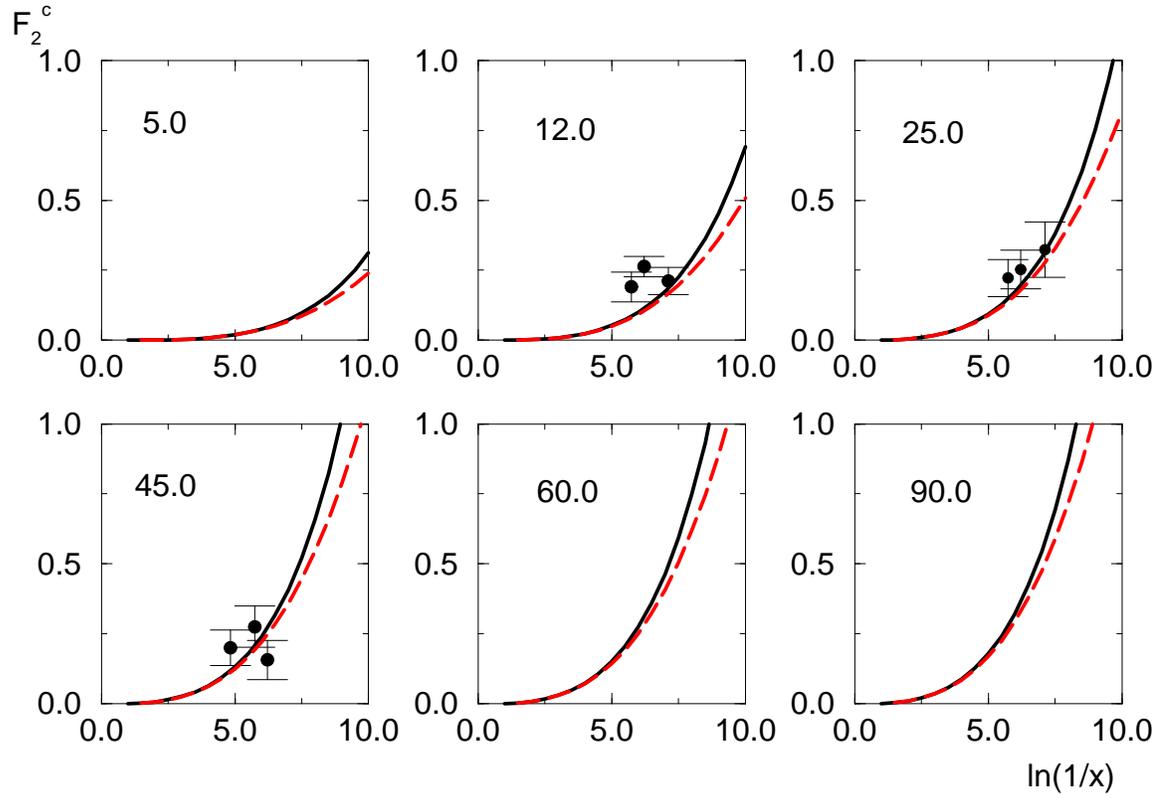,width=150mm}}
\caption{ The charm component of the structure function for different values of $Q^2$. 
 The solid curve represents  $F_2^c$  calculated using the GRV parameterization and the dashed curve as calculated using $xG$  obtained from the eikonal model. Data are from H1. }
\label{fig6}
\end{figure}

In  figure \ref{fig6} we present the behaviour of the $F_2^c$ structure 
function with (dashed curve) and without shadowing (solid curve) for different 
virtualities. We compare our results with the H1 data \cite{f2ch1}. 
Our predictions agree with the recent H1 data.
When compared with the GRV parameterization the shadowing corrections to 
$F_2^c$ present a suppression which increases with $ln(\frac{1}{x})$.
Therefore, new data, with better statistics, could isolate the shadowing 
corrections at HERA kinematic region.

\begin{figure}
\centerline{\psfig{file=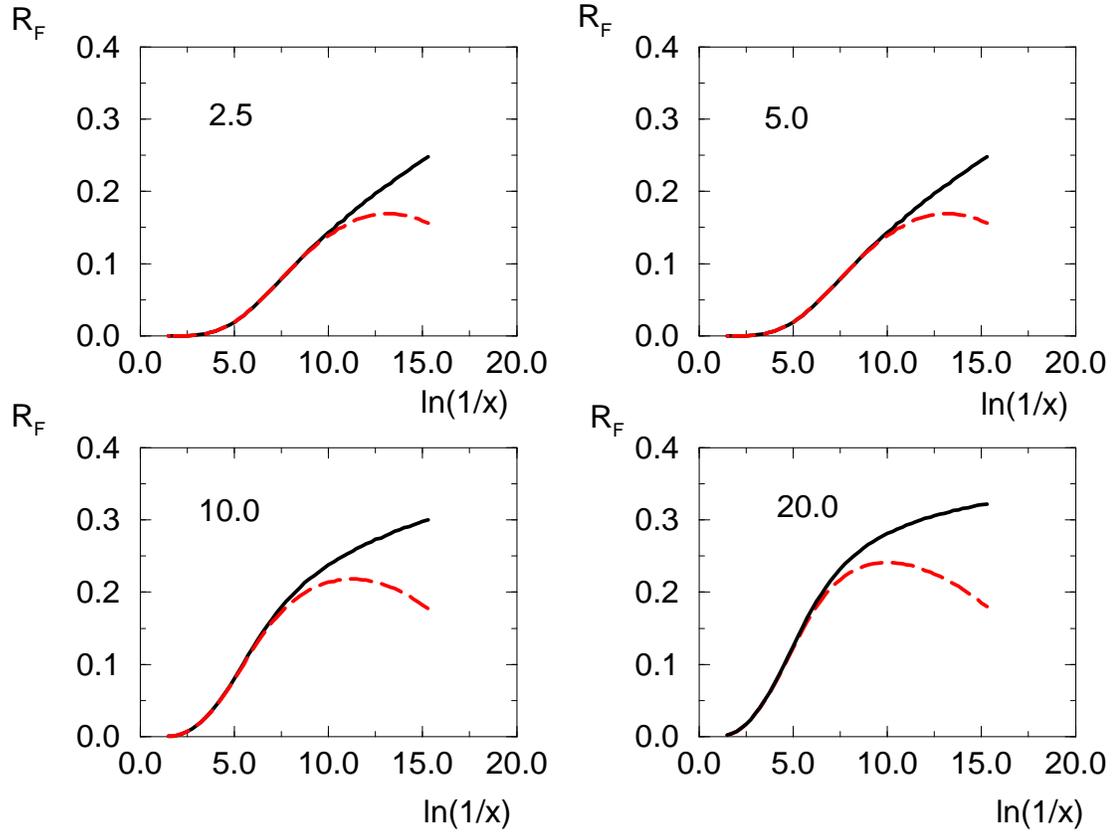,width=150mm}}
\caption{ The ratio $R_F= \frac{F_2^{c}}{F_2}$  as a function of the variable $ln(\frac{1}{x})$ for different virtualities.  The solid curve represents the predictions obtained  using the GRV parameterization and the dashed curve the predictions  obtained in the eikonal model. }
\label{fig7}
\end{figure}

As the shadowing corrections are distinct for the observables $F_2$ and $F_2^c$ we can estimate the modifications in the charm contribution to the $F_2$ at small $x$ by SC.
In  figure \ref{fig7}  we present our results for the ratio
\begin{eqnarray}
R_F= \frac{F_2^{c}(x,Q^2)}{F_2(x,Q^2)}\,\,.
\label{rf}
\end{eqnarray}
We can see that the behaviour of this ratio is strongly modified by the 
shadowing corrections. Therefore, we expect that the charm contribution do
not increase at small values of $x$ when compared with the predictions 
obtained using the usual parameterizations. Furthermore, we expect a strong 
modification in the charm production at small $x$. It is one of the main 
conclusions of this paper. This strong modification in the charm production shall occur in other associated observables, for instance in $J/\Psi$ production.
The shadowing corrections to the diffractive leptoproduction  of vector meson $J/\Psi$ was calculated in \cite{lmg}. The authors  obtained that the SC are large, but their results are affected by the uncertainty in the $J/\Psi$ wavefunction. Consequently, although the cross section of  diffractive leptoproduction  of $J/\Psi$ is proportional to the square of the gluon distribution, the accurate discrimination of SC in this process is still unlikely. 

Our conclusion of this section is that  the charm component $F_2^c$ is also a good observable to isolate the shadowing corrections. Therefore, new data, with better statistics will be very important in the determination of dynamics at HERA, and further applications.

\section{Conclusions}

The pQCD has furnished a remarkably successful framework to interpret a 
wide range 
of high energy lepton-lepton, lepton-hadron and hadron-hadron processes.
Through global analysis of these processes, detailed  information on the
parton structure of hadrons, especially the nucleon, has been obtained. 
The existing global analysis have been performed using the standard DGLAP evolution 
equations. However, in  the small $x$ region the DGLAP evolution equations are 
expected to breakdown, since new effects must occur in this kinematical  
region. One of these effects is the shadowing phenomenon. 

The recent $F_2$ 
data are well described in the framework of the DGLAP evolution equations 
with an appropriate choice of input distributions and the choice of the 
starting scale for the $Q^2$ evolution. Although no other ingredient has been 
needed to describe the data, we cannot  conclude that the shadowing corrections
are negligible at HERA kinematic region only from the analysis of $F_2$ data.
The proton structure function is inclusive on the behaviour of the gluon 
distribution.
In this paper we estimate the shadowing corrections to $F_L$ and $F_2^c$. 
The behaviour of these  observables is directly dependent on the behaviour 
of the gluon distribution and, therefore, strongly sensitive to the 
shadowing corrections. We shown that the SC to  $F_L$ and $F_2^c$
are much bigger than for the $F_2$ case. Our results are in accord with 
the recent few data. New data with better statistics are strongly 
welcome.

One of the main conclusions of this paper is the strong modification of 
charm component of the structure function. As the SC are different in the 
observables $F_2$ and $F_2^c$ the charm contribution do not increase at 
small $x$ as predicted by the usual parameterizations.

In this paper we estimate the shadowing corrections using an eikonal 
approach proposed in \cite{ayala2}.  The eikonal approach gives a 
sufficiently reliable result for the HERA kinematic region, however, 
a more accurate approach, to a  larger kinematic 
region than the HERA kinematic region, should consider as an input the solution 
of the generalized equation proposed in \cite{ayala2}.
In a future publication we will estimate the 
behaviour of some observables using the solution of the generalized equation.

The determination of the dynamics at small $x$  is fundamental to estimate the 
cross sections of the processes which will be studied in the future accelerators.
We expect that this paper could motivate a more accurate determination of $F_L$ 
and $F_2^c$ in the next years, since  the behaviour of these observables 
may explicitate the dynamics at small $x$ regime.

\section*{Acknowledgments}

This work was partially financed by CNPq and by Programa de Apoio a N\'ucleos de Excel\^encia (PRONEX), BRAZIL.

\newpage

\end{document}